\documentclass{ws-p8-50x6-00}

\usepackage{bm}

\begin{document}

\title{
Higher and Missing Resonances in \\ $\bm{\omega}$
photoproduction}

\author{Yongseok Oh}

\address{Institute of Physics and Applied Physics,
Yonsei University, Seoul 120-749, Korea
\\E-mail: yoh@phya.yonsei.ac.kr}

\author{Alexander I. Titov}

\address{Bogoliubov Laboratory of Theoretical Physics, JINR,
Dubna 141980, Russia\\
E-mail: atitov@thsun1.jinr.ru}

\author{T.-S. H. Lee}

\address{Physics Division, Argonne National Laboratory, Argonne,
Illinois 60439, U.S.A. \\
E-mail: lee@theory.phy.anl.gov}


\maketitle

\abstracts{
We study the role of the nucleon resonances ($N^*$) in $\omega$
photoproduction by using the quark model resonance parameters
predicted by Capstick and Roberts.
The employed $\gamma N \to N^*$ and $N^* \to \omega N$ amplitudes
include the configuration mixing effects due to the residual
quark-quark interactions.
The contributions from the nucleon resonances are found to be
important in the differential cross sections at large scattering
angles and various spin observables.
In particular, the parity asymmetry and beam-target double asymmetry
at forward scattering angles are suggested for a crucial test of
our predictions.
The dominant contributions are found to be from $N\frac32^+ (1910)$,
a missing resonance, and $N\frac32^- (1960)$ which is identified as
the $D_{13}(2080)$ of the Particle Data Group.
}

The nucleon resonances predicted by the constituent quark models have a
much richer spectrum than what have been observed in pion-nucleon
scattering\cite{IK77-80}.
The origin of this ``missing resonance problem'' has been ascribed to
the possibility that many predicted nucleon resonances ($N^*$) could
couple weakly to the $\pi N$ channel\cite{CR00}.
Therefore it would be legitimate to study the nucleon resonances
in other reactions and vector meson electromagnetic production is one of
them which are under investigation at current experimental facilities
such as TJNAF, ELSA-SAPHIR, GRAAL, and LEPS of SPring-8.
Theoretically the role of the nucleon resonances was studied by Zhao
{\em et~al.\/}\cite{ZLB98,Zhao01} based on an effective Lagrangian
method within the SU(6) $\times$ O(3) constituent quark model.

Our study on the nucleon resonances in vector meson
photoproduction\cite{OTL01} is based on the quark model predictions
by Capstick and Roberts\cite{CR94}, where the configuration
mixing effects due to the residual quark-quark interactions are included
and the hadron decays are calculated by using the ${}^3P_0$ model.
The predicted baryon wave functions and the $N^*$ decay amplitudes are
considerably different from those of the SU(6) $\times$ O(3) quark
model\cite{ZLB98}.
Thus it would be interesting to find the differences in the model
predictions on vector meson photoproduction, which can be tested
experimentally.

In this work we focus on $\omega$ photoproduction\cite{OTL01} since
its non-resonant reaction mechanisms are rather well understood and
the isosinglet nature of the $\omega$ meson allows the contributions
from the isospin-$1/2$ nucleon resonances only.
Earlier studies on $\omega$ photoproduction\cite{OLD} show that the
reaction is dominated by diffractive processes at high energies, i.e.,
via the Pomeron exchange, and by one-pion exchange at low energies.
It is therefore reasonable to follow the earlier theoretical analyses
and assume that the non-resonant amplitude of $\omega$ photoproduction
can be calculated from these two well-established mechanisms with some
refinements.
The resulting model then can be a starting point for investigating
the $N^*$ effects.

\begin{figure}[t]
\centering
\epsfig{file=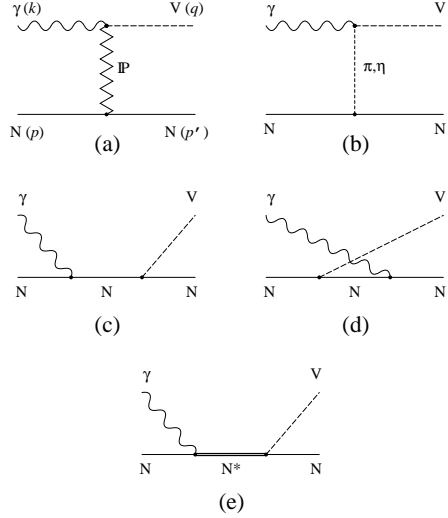, width=6cm}
\caption{
Diagrammatic representation of $\omega$ photoproduction mechanisms:
(a) Pomeron exchange, (b) ($\pi,\eta$) exchange, (c) direct nucleon
term, (d) crossed nucleon term, and (e) $s$-channel nucleon
excitations.}
\label{fig:diag}
\end{figure}

Our model for $\omega$ photoproduction, therefore, can be described by
the diagrams shown in Fig.~\ref{fig:diag}.
The Pomeron exchange [Fig.~\ref{fig:diag}(a)] is known to govern the
total cross sections and differential cross sections at low $|t|$ in the
high energy region for electromagnetic production of vector mesons.
For this model, we follow the Donnachie-Landshoff model\cite{DL84-92}.
The details on this model can be found, e.g., in
Refs.~\cite{LM95,PL97} and the resulting amplitude and the parameters
are summarized in Refs.~\cite{OTL01,OTL00}.

The pseudoscalar meson exchange amplitudes [Fig.~\ref{fig:diag}(b)] are
calculated from the following effective Lagrangians:
\begin{eqnarray}
{\cal L}_{\omega \gamma \varphi}^{} &=&
\frac{e g_{\omega\gamma \varphi}}{M_V}
\epsilon^{\mu\nu\alpha\beta}
\partial_\mu \omega_\nu \partial_\alpha A_\beta\, \varphi,\qquad
\nonumber\\
{\cal L}_{\varphi NN}^{} &=&
-i g_{\pi NN} \bar N \gamma_5\tau_3 N \pi^0
-i g_{\eta NN} \bar N \gamma_5 N \eta,
\end{eqnarray}
where $\varphi = (\pi^0,\eta)$ and $A_\beta$ is the photon field.
We use $g_{\pi NN}^2/4\pi = 14$ and $g^2_{\eta NN}/4\pi = 0.99$ for the
$\pi NN$ and $\eta NN$ coupling constants, respectively.
The coupling constants $g_{\omega\gamma\varphi}$ can be estimated
through the decay widths of $\omega \to \gamma\pi$ and
$\omega \to \gamma \eta$ \cite{PDG00} which lead to
$g_{\omega\gamma\pi} = 1.823$ and $g_{\omega\gamma\eta} = 0.416$.
The higher mass of the $\eta$ and the associated small coupling
constants suppress the $\eta$ exchange contribution compared with
the $\pi$ exchange.
The $\varphi NN$ and $\omega\gamma\varphi$ vertices are dressed by
the form factors,
\begin{equation}
F_{\varphi NN}^{} (t) = \frac{\Lambda_\varphi^2 - M^2_\varphi}
{\Lambda_\varphi^2 -t},  \qquad
F_{\omega\gamma\varphi}^{} (t) =
\frac{\Lambda_{\omega\gamma\varphi}^2-M_\varphi^2}
{\Lambda_{\omega\gamma\varphi}^2-t} ,
\label{PS:FF}
\end{equation}
with $\Lambda_\pi = 0.6$ GeV and $\Lambda_{\omega\gamma\pi} = 0.7$
GeV \cite{OTL01}, $\Lambda_\eta = 1.0$ GeV and
$\Lambda_{\omega\gamma\eta} = 0.9$ GeV \cite{TLTS99}.

The nucleon pole terms [Fig.~\ref{fig:diag}(c,d)] are calculated from
the following interaction Lagrangians:
\begin{eqnarray}
{\cal L}_{\gamma NN}^{} & = &
- e \bar{N} \left( \gamma_\mu \frac{1+\tau_3}{2} {A}^\mu
- \frac{\kappa_N^{}}{2M_N^{}} \sigma^{\mu\nu} \partial_\nu A_\mu \right)
  N,
\nonumber\\
{\cal L}_{\omega NN}^{} & = &
- g_{\omega NN}^{} \bar{N} \left( \gamma_\mu {\omega}^\mu
- \frac{\kappa_\omega}{2M_N^{}} \sigma^{\mu\nu}
\partial_\nu \omega_\mu \right) N,
\end{eqnarray}
with the anomalous magnetic moment of the nucleon $\kappa_{p(n)} =
1.79$ $(-1.91)$.
For the coupling constants we use $g_{\omega NN}^{} = 10.35$ and
$\kappa_\omega^{} = 0$ \cite{SL96,RSY99}.
In order to dress the vertices, we include the form
factor\cite{HBMF98a},
\begin{eqnarray}
F_N (r) = \frac{\Lambda_N^4}{\Lambda_N^4  - (r - M_N^2)^2},
\label{N:FF}
\end{eqnarray}
where $r = s$ or $t$ and $\Lambda_N = 0.5$ GeV.

\begin{figure}[t]
\centering
\epsfig{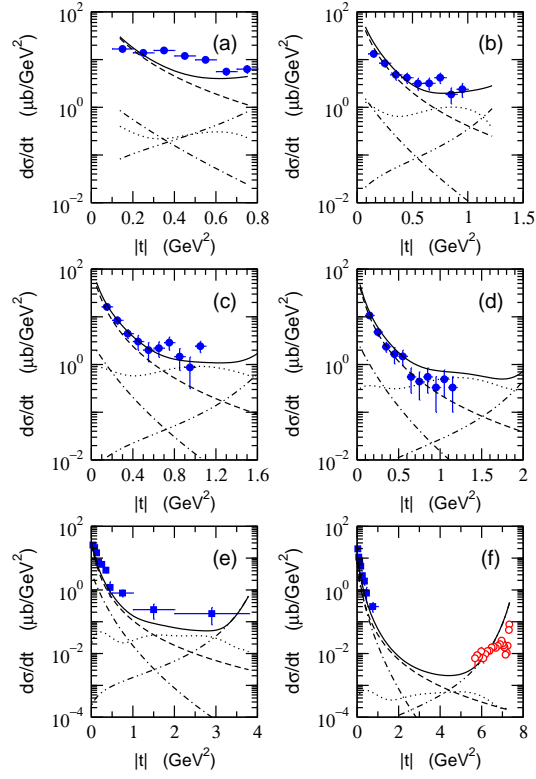}
\caption{
Differential cross sections for $\gamma p\to p\omega$ reaction
as a function of $|t|$ at $E_\gamma =$ (a) $1.23$, (b) $1.45$, (c)
$1.68$, (d) $1.92$, (e) $2.8$, and (f) $4.7$ GeV.
The results are from pseudoscalar-meson exchange (dashed),
Pomeron exchange (dot-dashed), direct and crossed nucleon terms
(dot-dot-dashed), $N^*$ excitation (dotted), and the full amplitude
(solid).
Data are taken from Ref.~\protect\cite{Klein96-98} [filled circles in
(a,b,c,d)],
Ref.~\protect\cite{BCEK73} [filled squares in (e,f)], and
Ref.~\protect\cite{CDGL77} [open circles in (f)].}
\label{fig:dsdt}
\end{figure}

With the non-resonant amplitudes discussed above we estimate the nucleon
resonance contributions by making use of the quark model
predictions\cite{CR94} on the resonance photoexcitation $\gamma N \to N^*$
and the resonance decay $N^* \to N \omega$.
In this work we consider the $s$-channel diagrams shown in
Fig.~\ref{fig:diag}(e).
The crossed diagrams cannot be calculated from the informations
available in Ref.~\cite{CR94}.
The resonant amplitude in the center of mass frame is written as
\begin{eqnarray}
I^{N^*}_{m_f,m_\omega,m_i,\lambda_\gamma}({\bf q},{\bf k})
&=& \sum_{J,M_J^{}}
\frac{1}
{\sqrt{s} - M_R^J + \frac{i}{2}\Gamma^J(s)}
{\cal M}_{N^*\to N'\omega}({\bf q};m_f^{},m_\omega^{};J,M_J^{})
\nonumber \\ && \quad \mbox{} \times
{\cal M}_{\gamma N \to N^*}({\bf k};m_i^{},\lambda_\gamma^{};J,M_J^{}),
\label{T:N*}
\end{eqnarray}
where $M^J_R$ is the $N^*$ mass of spin quantum numbers
$(J, M_J)$, and $m_i$ , $m_f$, $\lambda_\gamma$, and $m_\omega$ are the
spin projections of the initial nucleon, final nucleon, incoming photon,
and outgoing $\omega$ meson, respectively.
In this study, we consider 12 positive parity and 10 negative parity
nucleon resonances up to spin-9/2.
The explicit form of the resonant amplitude and the details on the
calculations can be found in Ref.~\cite{OTL01} as well as the considered
nucleon resonances and their parameters.

Our results for the differential cross section are shown in
Fig.~\ref{fig:dsdt}, which shows that
the data can be described to a very large extent in the
considered energy region, $ E_\gamma \leq 5 $ GeV.
It is clear that the contributions due to the $N^*$ excitations (dotted
curves) and the direct and crossed nucleon terms (dot-dot-dashed curves)
help bring the agreement with the data at large angles.
The forward angle cross sections are mainly due to the interplay between
the pseudoscalar-meson exchange (dashed curves) and the Pomeron exchange
(dot-dashed curves).
The main problem here is in reproducing the data at $E_\gamma = 1.23$
GeV.
This perhaps indicates that the off-shell contributions from $N^*$'s
below $\omega N$ threshold are important at very low energies.

\begin{figure}[t]
\centering
\epsfig{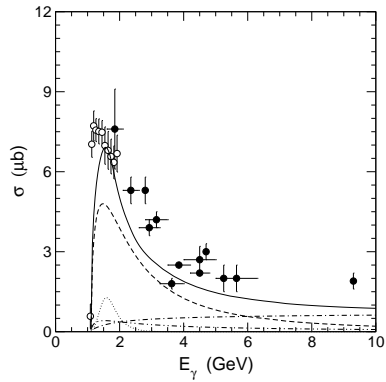}
\caption{
Total cross section of $\omega$ photoproduction.
The notations are the same as in Fig.~\protect\ref{fig:dsdt}.
Data are taken from Refs.~\protect\cite{Klein96-98,BCEK73,Durham}.}
\label{fig:totcs}
\end{figure}

The contribution from the nucleon resonances to total cross sections is
shown in Fig.~\ref{fig:totcs}.
To have a better understanding of the resonance contributions, we
compare the contributions from the considered $N^*$'s to the
differential and total cross sections.
We found that the contributions from $N\frac32^+ (1910)$ and
$N\frac32^- (1960)$ are the largest at all energies up to $E_\gamma = 3$
GeV.
The $N\frac32^+ (1910)$ is a missing resonance, while $N\frac32^- (1960)$
is identified as a two star resonance $D_{13}(2080)$ of the Particle
Data Group\cite{PDG00}.
This result is significantly different from the quark model calculations
of Ref.~\cite{ZLB98}.
The difference between the two calculations is not surprising
since the employed quark models are rather different.
In particular, our predictions include the configuration mixing effects
due to residual quark-quark interactions.
The discrepancy of our prediction with the experimental data at very low
energy is again expected to be due to the nucleon resonances below
$\omega N$ threshold, which are missed in our calculation.

Instead of the cross sections, the polarization
asymmetries\cite{TOYM98} provide more appropriate tools to investigate
the role of the nucleon resonances in $\omega$ photoproduction\cite{OTL01}.
We first examined the single spin asymmetries\cite{OTL01}.
Although our predictions are significantly different from those of
Ref.~\cite{ZLB98}, we confirm their conclusion that those asymmetries
are sensitive to the nucleon resonances but mostly at large $|t|$
region.

\begin{figure}[t]
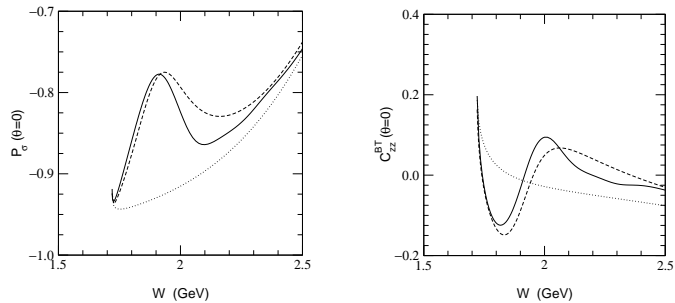

\centering
\epsfig{file=fig4a.eps, width=4cm} \qquad
\epsfig{file=fig4b.eps, width=4cm}
\caption{
Parity asymmetry $P_\sigma$ and beam-target double asymmetry $C^{\rm
BT}_{zz}$ at $\theta=0$ as functions of $W$.
The dotted curves are calculated without including $N^*$ effects, the
dashed curves include contributions of $N\frac32^+(1910)$ and
$N\frac32^-(1960)$ only, and the solid curves are calculated with all
$N^*$'s up to spin-9/2.}
\label{fig:asym}
\end{figure}

More clear signal for the nucleon resonances can be found from the
parity asymmetry (or photon polarization asymmetry) and the
beam-target double asymmetry.
These asymmetries are sensitive to the $N^*$ contributions at forward
angles, where precise measurements might be more favorable because the
cross sections are peaked at $\theta = 0$.
The parity asymmetry is defined as\cite{SSW}
\begin{equation}
P_\sigma =
\frac{d\sigma^N - d\sigma^U}{d\sigma^N + d\sigma^U}
= 2\rho^1_{1-1}-\rho^1_{00},
\end{equation}
where $\sigma^N$ and $\sigma^U$ are the cross sections due to the
natural and unnatural parity exchanges respectively, and
$\rho^{i}_{\lambda,\lambda^\prime}$ are the vector-meson spin
density matrices.
The beam-target double asymmetry is defined as\cite{TOYM98}
\begin{equation}
C^{BT}_{zz} =
\frac{d\sigma(\uparrow\downarrow) - d\sigma(\uparrow\uparrow)}
{d\sigma(\uparrow\downarrow) + d\sigma(\uparrow\uparrow)},
\end{equation}
where the arrows represent the helicities of the incoming photon and
the target proton.
In Fig.~\ref{fig:asym}, we show the results from calculations with
(solid curves) and without (dotted curves) including $N^*$ contributions.
The difference between them is striking and can be unambiguously tested
experimentally.
Here we also find that the $N\frac32^+ (1910)$ and $N\frac32^- (1960)$
are dominant.
By keeping only these two resonances in calculating the resonant part of
the amplitude, we obtain dashed curves which are not too different from
the full calculations (solid curves).

To summarize, we investigated the role of the nucleon resonances in
$\omega$ photoproduction.
We found that the inclusion of the resonance amplitudes leads to a
better description of the observed total and differential cross sections.
It is also found that the $N\frac32^+ (1910)$ and $N\frac32^- (1960)$
are dominant in the resonance amplitudes.
As a further study on the nucleon resonances, we suggest to measure the
parity asymmetry and beam-target double asymmetry.
Experimental test of our predictions will be a useful step toward
resolving the so-called ``missing resonance problem'' or distinguishing
different quark model predictions and could be done at current
electron/photon facilities.
Theoretically the predictions should be improved further.
For example, the form factor of the vertices including the nucleon
resonances should be studied in detail in the given quark models
and the crossed $N^*$ terms as well as the nucleon resonances below
$\omega N$ threshold should be studied.
Finally, the effects due to the initial and final state interactions
must be also investigated especially at low energies.

\section*{Acknowledgments}
This work was supported in part by the Brain Korea 21 project of Korean
Ministry of Education, the International Collaboration Program of
KOSEF under Grant No. 20006-111-01-2, Russian Foundation for Basic
Research under Grant No. 96-15-96426, and U.S. DOE Nuclear Physics
Division Contract No. W-31-109-ENG-38.


\end{document}